\documentstyle[12pt]{article}

\newlength{\myleftmargin}
\newlength{\paperwidth}
\begin{document}
\renewcommand{\thefootnote}{\fnsymbol{footnote}}
\begin{flushright}
KOBE--FHD--94--02\\
February~~24~~~1994
\end{flushright}

\baselineskip=20pt

\begin{center}
{\large \bf How to probe the spin--dependent gluon
distributions\footnote[2]{Talk presented at meeting on ``Dynamics of
Quarks $\cdot$ Hadrons" (Yukawa Inst. for Theor. Phys., Kyoto,
Dec. 6 -- 8, 1993).}}\\

\vspace{2em}

T. Morii$^{1, 2}$, S. Tanaka$^1$ and T. Yamanishi$^2$\\

\vspace{1em}

Faculty of Human Development, Division of Sciences for Natural Environment,\\
Kobe University, Nada, Kobe 657, Japan$^1$\\
Graduate School of Science and Technology,
Kobe University, Nada, Kobe 657, Japan$^2$\\

\vspace{4em}

{\bf Abstract}
\end{center}

Two--spin asymmetries $A_{LL}^{\pi^0}$ are calculated for various types of
spin--dependent gluon distributions. It is concluded that the E581/704 data
on $A_{LL}^{\pi^0}$ do not necessarily rule out the large gluon polarization
but restrict severely the $x$ dependence of its distribution. Moreover,
$A_{LL}^{J/\psi}$ are calculated for the forthcoming test of spin--dependent
gluon distributions.

\baselineskip=20pt

\vspace{5em}

The advent of {\it so--called} ``the proton spin crisis'' which has emerged
from the measurement of $g_1^p(x)$ by the EMC Collaboration, has stimulated a
great theoretical and experimental activity in paricle physics \cite{EMC}.
So far various theoretical approaches have been provided to get rid of the
crisis \cite{crisis}.
Although some of them are very successful, a lot of problems remain to be
solved.
One of the remaining problems is on the polarized gluon distributions
($\delta G(x)$) in a proton. In this Talk, we are concentrated on several
physical processes such as inclusive $\pi^0$-- \cite{Rams91}, high $p_T$ direct
photon-- \cite{Cheng90} and inclusive J/$\psi$--productions \cite{Donch90}
in polarized hadron--polarized hadron reactions,
which possess important informations on the polarized gluons.
Here we discuss only $\pi^0$-- and J/$\psi$--productions.
The interesting physical parameter to be discussed is the two--spin asymmetry
$A_{LL}$ as a function of transverse momenta $p_T$ of produced particles like
$\pi^0$, $\gamma$ and J/$\psi$. $A_{LL}$ is defined as
\begin{eqnarray}
A_{LL}~&=&~\frac{\left[d\sigma_{\uparrow\uparrow}-d\sigma_{\uparrow\downarrow}+
d\sigma_{\downarrow\downarrow}-d\sigma_{\downarrow\uparrow}\right]}
{\left[d\sigma_{\uparrow\uparrow}+d\sigma_{\uparrow\downarrow}+
d\sigma_{\downarrow\downarrow}+d\sigma_{\downarrow\uparrow}\right]}
\nonumber\\
&=&~\frac{Ed\Delta\sigma/d^3p}{Ed\sigma/d^3p}~,
\label{eqn:dfnALL}
\end{eqnarray}
where $d\sigma_{\uparrow\downarrow}$, for instance, denotes that the helicity
of a beam particle is positive and that of a target particle is negative.

In order to investigate how these processes are affected by spin--dependent
gluon distributions, we take the following types of $x\delta G(x)$ :
\begin{enumerate}
\renewcommand{\labelenumi}{(\alph{enumi})}
\item our model \cite{Morii}~;
\begin{eqnarray}
&&x\delta G(x, Q^2=10.7{\rm GeV}^2)=3.1x^{0.1}(1-x)^{17}
{}~{\rm then}~~~~~~~~~~~~
\label{eqn:typeA}\\
&&\Delta G(Q^2_{EMC})=6.32~.\nonumber
\end{eqnarray}
\item Cheng--Lai type model \cite{HYCheng}~;
\begin{eqnarray}
&&x\delta G(x, Q^2=10{\rm GeV}^2)=3.34x^{0.31}(1-x)^{5.06}(1-0.177x)
{}~{\rm then}
\label{eqn:typeB}\\
&&\Delta G(Q^2_{EMC})=5.64~.\nonumber
\end{eqnarray}
\item BBS model \cite{BBS}~;
\begin{eqnarray}
&&x\delta G(x, Q^2=4{\rm GeV}^2)=0.281\left\{(1-x)^4-(1-x)^6\right\}+
1.1739\left\{(1-x)^5-(1-x)^7\right\}
\label{eqn:typeC}\\
&&{\rm then}~~~\Delta G(Q^2_{EMC})=0.53~.\nonumber
\end{eqnarray}
\item no gluon polarization model \cite{HYCheng}~;
\begin{equation}
x\delta G(x, Q^2=10{\rm GeV}^2)=0~~{\rm then}~~\Delta G(Q^2_{EMC})=0~.
\label{eqn:typeD}
\end{equation}
\end{enumerate}
Among these distributions, $\Delta G$ of types (a) and (b) are large while
those of types (c) and (d) are small and zero, respectively.
The $x$ dependence of these
distributions are depicted in Fig.1, where $x\delta G(x, Q^2)$ of types (b),
(c) and (d) are evolved up to $Q^2=10.7$GeV$^2$ by the Altarelli--Parisi
equations.
The $x\delta G$ which is taken up so far by most of people
\cite{gluons} has almost the same behavior as that of type (b) and has
large $\Delta G$.
As can be seen, the $x\delta G(x)$ of type (b) has a peak at $x\approx 0.05$
and gradually decreases with increasing $x$ while that of (a) has a sharp
peak at $x<0.01$ and rapidly decreases with $x$.

First, we discuss the inclusive $\pi^0$--production.
So far, only this process has been measured
by the E581/704 Collaboration at Fermilab\cite{E581} by using longitudinally
polarized proton (antiproton) beams and longitudinally polarized proton
targets. Using the spin--dependent gluon distribution functions
((a)$\sim$(d)) presented above, we have calculated $A_{LL}^{\pi^0}(pp)$ and
$A_{LL}^{\pi^0}(\overline{p}p)$, which are shown in Figs.2 and 3 for
$\sqrt s=20$ GeV and $\theta =90^{\circ}$, respectively.
Here we typically choose $Q^2=4p_T^2$ with the transverse momentum $p_T$ of
$\pi^0$.

Comparing theoretical predictions with the experimental data, we see that not
only the no gluon polarization model (type (d)) but also our model (type (a))
seem to be consistent with the experimental data for both
$pp$ and $\overline{p}p$ collisions. It is remarkable to see that type (a)
works well though it has large $\Delta G$.
Owing to the kinematical constraint of $x$ in the hard--scattering parton
model, the contributions from $0<x<0.05$
to $A^{\pi^0}_{LL}(\stackrel{\scriptscriptstyle(-)}{p}\stackrel{}{p})$ are
vanishing. Accordingly, there are no significant contributions from the
spin--dependent gluon distribution of type (a) to $A_{LL}^{\pi^0}$ though
$\Delta G(Q^2)$ for this case is quite large.
However, if we take the polarized gluon distribution $x\delta G(x)$ of
type (b) which is still large for $x>0.05$, we have a significant contribution
from the large $x\delta G(x)$ to $A_{LL}^{\pi^0}$ and then the result
becomes inconsistent with the E581/704 data. Furthermore, if the value of
$x\delta G(x)$ is not very small for $x>0.15$ even though $\Delta G(x)$ is
small (as in the case of type (c)), the calculation does not agree with the
experimental data. Therefore, one can conclude that a
large gluon polarization inside a proton is not necessarily ruled out
but the shape of the spin--dependent gluon distribution function is
strongly constrained by the E581/704 data.

Next, in order to get a more direct information of spin--dependent gluon
distributions, let us discuss the inclusive J/$\psi$ production process in
polarized proton--polarized proton collisions \cite{Donch90}. Since the
J/$\psi$ productions come out only via gluon--gluon fusion processes at the
lowest order of QCD diagrams, this quantity is strongly sensitive to the
spin--dependent gluon distribution in a proton.
For estimation of $A_{LL}^{J/\psi}(pp)$, we take the
spin--dependent gluon distributions (a), (b), (c) and (d) given by
eqs.(\ref{eqn:typeA}), (\ref{eqn:typeB}), (\ref{eqn:typeC}) and
(\ref{eqn:typeD}).
Setting $\theta=90^{\circ}$ ($\theta$ is the production angle of J/$\psi$ in
the CMS of colliding protons) and using the spin--independent gluon
distribution function of the DO parametrization \cite{Duke} for (a),
the DFLM \cite{DFLM} for (b) and (d) and the BBS \cite{BBS} for (c), we have
calculated $A_{LL}^{J/\psi}(pp)$ for some choices of $Q^2$ ;
$Q^2=m^2_{J/\psi}+p_T^2$, $4p_T^2$, $(\hat s\hat t\hat u)^{1/3}$, $-\hat t$
and so on.
We see that $A_{LL}^{J/\psi}(pp)$ for each type of the spin--dependent
gluon distributions is insensitive to the choice of $Q^2$.
Thus, we here take $Q^2=m_{J/\psi}^2+p_T^2$ by taking the mass effect of the
J/$\psi$ particle into account. The results of $A_{LL}^{J/\psi}(pp)$ are
shown in Fig.4 as a function of $p_T$ of the J/$\psi$ at (A) $\sqrt s=20$ and
(B) $100$ GeV. At $\sqrt s=20$ GeV our largely polarized gluon distribution,
(a), contributes little to
$A_{LL}^{J/\psi}(pp)$ in all $p_T$ regions because the kinematical
region near the peak of $x\delta G(x)$ is truncated.
The $A_{LL}^{J/\psi}$ predicted with type (a) is not so significantly
different from that with no gluon polarization (type (d)), and hence we cannot
practically find the difference between them. However, for higher energies such
as $\sqrt s=100$ GeV, we can distinguish types (a) from (d) for
spin--dependent gluon distributions by choosing a moderate $p_T$ region.
In addition, one can see that the behavior of $A_{LL}^{J/\psi}$ for types (b)
and (c) largely differs from that for
types (a) and (d) at $\sqrt s=20$ and $100$ GeV. Therefore, it is expected
that one can either rule out or confirm types (b) and (c) by measuring
$A_{LL}^{J/\psi}$ particularly in rather large $p_T$ regions.

In summary, we have studied the effect of the polarized gluon distributions on
the two--spin asymmetry $A_{LL}^{\pi^0}$ and $A_{LL}^{J/\psi}$ in the
polarized proton (antiproton)--polarized proton collisions. We have concluded
that although the E581/704 data of $A_{LL}^{\pi^0}$ are very useful to
examine the behavior of polarized gluon distributions, they are not enough
to distinguish type (a) from type (d). To get more deep understandings of
polarized gluons in a proton, we have to analyze other reactions such as
polarized $\ell$p collisions, which are now under investigation.

\vfill\eject

\begin{center}
{\large \bf Figure captions}
\end{center}
\begin{description}
\item[Fig. 1:] The $x$ dependence of $x\delta G(x, Q^2)$ for various types
(a)--(d) given by eqs.(\ref{eqn:typeA})-- (\ref{eqn:typeD})
at $Q^2=10.7$ GeV$^2$.

\vspace{2em}

\item[Fig. 2:] Two--spin asymmetry $A_{LL}^{\pi^0}(pp)$ for $\sqrt s=20$ GeV
and $\theta=90^{\circ}$, calculated with various types of
$x\delta G(x)$, as a function of transverse momenta $p_T$ of $\pi^0$.
The solid, dashed, small--dashed and dash--dotted lines indicate the results
using
types (a), (b), (c) and (d) in eqs.(\ref{eqn:typeA}), (\ref{eqn:typeB}),
(\ref{eqn:typeC}) and (\ref{eqn:typeD}), respectively. Experimental data are
taken from \cite{E581}.

\vspace{2em}

\item[Fig. 3:] Two--spin asymmetry $A_{LL}^{\pi^0}(\overline{p}p)$ for
$\sqrt s=20$ GeV and $\theta=90^{\circ}$, calculated with types (a)--(d)
for $x\delta G(x)$, as a function of transverse
momenta $p_T$ of $\pi^0$. Data are taken from \cite{E581}.

\vspace{2em}

\item[Fig. 4:] Two--spin asymmetries $A_{LL}^{J/\psi}(pp)$ for
$\theta=90^{\circ}$ calculated with types (a)--(d) for
$x\delta G$, as a function of transverse momenta $p_T$ of J/$\psi$ at (A)
$\sqrt s=20$ GeV, and (B) $\sqrt s=100$ GeV. The solid, dashed, small--dashed
and dash--dotted curves correspond to types (a), (b), (c) and (d),
respectively. $Q^2$ is typically taken to be $m^2_{J/\psi}+p_T^2$.
\end{description}
\end{document}